# Exciton-trion dynamics of a single molecule in a radio-frequency cavity

Jiří Doležal[1,2], Sofia Canola[1], Pablo Merino[3,4]*, Martin Švec[1,5]*

[1]Institute of Physics, Czech Academy of Sciences, Cukrovarnická 10/112, CZ16200 Praha 6, Czech Republic

[2]Faculty of Mathematics and Physics, Charles University, Ke Karlovu 3, CZ12116 Praha 2, Czech Republic

[3]Instituto de Ciencia de Materiales de Madrid, CSIC, Sor Juana Inés de la Cruz 3, E28049 Madrid, Spain

[4]Instituto de Física Fundamental, CSIC, Serrano 121, E28006 Madrid, Spain

[5]Regional Centre of Advanced Technologies and Materials, Šlechtitelů 27, CZ78371 Olomouc, Czech Republic

* pablo.merino@csic.es, svec@fzu.cz

Charged optical excitations (trions) generated by charge carrier injection are crucial for emerging optoelectronic technologies as they can be produced and manipulated by electric fields. Trions and neutral excitons can be efficiently induced in single molecules by means of tip-enhanced spectromicroscopic techniques. However, little is known of the exciton-trion dynamics at single molecule level as this requires methods permitting simultaneous sub-nanometer and sub-nanosecond characterization. Here, we investigate exciton-trion dynamics by phase fluorometry, combining radio-frequency modulated scanning tunnelling luminescence with time-resolved single photon detection. We generate excitons and trions in single Zinc Phthalocyanine (ZnPc) molecules on NaCl/Ag(111), determine their dynamics and trace the evolution of the system in the picosecond range with atomic resolution. In addition, we explore dependence of effective lifetimes on bias voltage and propose a conversion of neutral excitons into trions via charge capture as the primary mechanism of trion formation.

The dynamics of optical excitations provides insights into the photophysics of many-body quantum states of single molecules. *(1)* It is also a key to develop efficient single photon quantum cryptography and quantum computing protocols. *(2-4)* Although most experiments rely on ensemble or bulk measurements, excitation and control of electron-hole bound states in single molecules and defects in molecular solids are possible for diluted emitters. *(5,6)* Recent developments in tip-enhanced spectroscopies, prominently in scanning tunneling microscopy luminescence (STML) opened an atomic-scale window to explore the mechanisms generating singlet, *(7,8)* triplet *(9,10)* and doublet *(11)* excitons in single molecules and the role of their nanoscopic environment. Emission from positively and negatively charged excitons (trions) was recently discovered for single Zinc and Platinum Phthalocyanine (ZnPc, PtPc) emitters, respectively. *(12,13)*. Despite important advances in measuring fast dynamics of optical excitations on the nanoscale by application of Hanbury Brown Twiss (HBT) interferometry *(14-19)* and time-resolved STML, *(20-22)* capturing the combined dynamics of excitons and trions in single molecules remained a challenge.

In order to access the trion and exciton dynamics in a single-molecule, we devised phase fluorometry scheme, combining radio-frequency (RF) electrical modulation and picosecond single-photon counting detection within a scanning probe microscopy setup equipped with optical path (scheme in Fig. 1A). Our radio-frequency phase-shift (RF-PS) technique is based on evaluation of the phase difference between a harmonic electrical modulation of a given frequency applied at the system and the delayed optical response at particular wavelength, arising due to a finite decay rate of excitons within the molecule. The reference and delayed responses are measured as histograms of photon arrival times of a plasmonic signal from the substrate and from the molecule, respectively. The characteristic radiative lifetime ($\tau$) of an exciton in the tunnel junction is determined from the phase difference ($\Delta\varphi$) for a given driving frequency $f$ using the relation $\tau =\tan(\Delta\varphi)/2\pi f$. *(23)*

By means of the RF-PS method we measure the dynamics of singlet excitons and trions on the ZnPc on 3 layers of NaCl on Ag(111). Injecting holes and electrons into a single ZnPc induces optical excitations that can decay radiatively. The emission is enhanced by the increased optical density in the picocavity. *(24)* The visible/near-infrared electroluminescence spectra of ZnPc (Fig. 1B) show peaks corresponding to the emission of neutral (X) at 1.89 eV and positively charged excitons ($X^+$) at 1.52 eV. *(12)* For each line, we measured the phase shift with respect to the reference and determined the exciton effective lifetimes as 635(51) ps for X and 348(55) ps for $X^+$, at bias voltage ($V_{DC}$) of -2.2 V and -3.2 V, respectively. The lifetime measured for X is in excellent agreement with previous HBT measurements;*(16,25)* no determination of the lifetime of $X^+$ is found in the literature as yet.

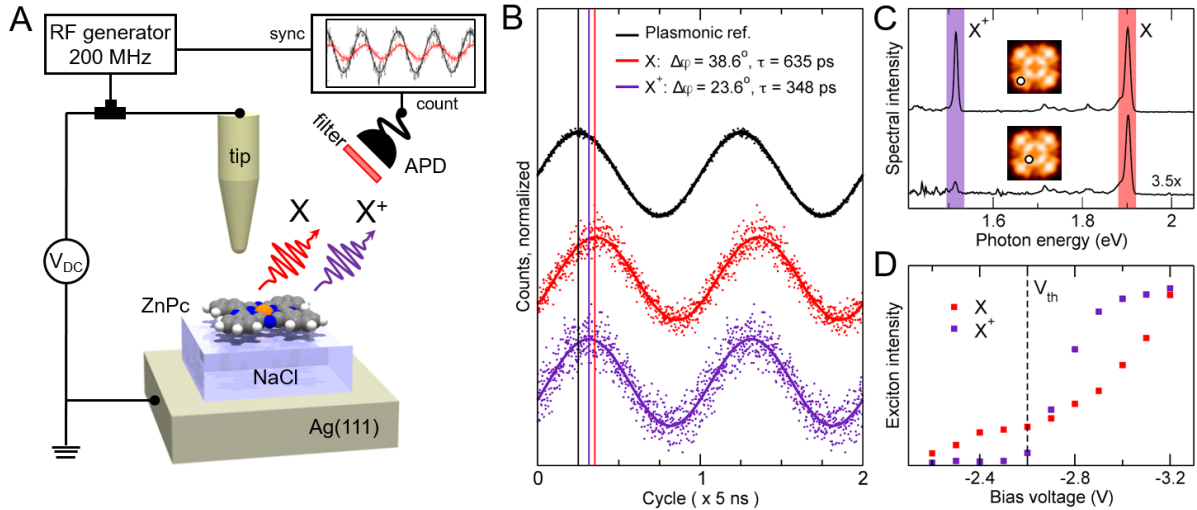

**Figure 1. Radio-frequency phase shift method in STML applied to a single molecule.** (**A**) Schematic representation of the experimental setup. A single ZnPc molecule on 3 layers of NaCl is inspected by STML. The tunnelling bias voltage ($V_{DC}$) is modulated with a harmonic signal at 200 MHz ($V_{AC}$). The photon arrival times are recorded using a single photon detector and counter synchronized with the radiofrequency generator. (**B**) Radio-frequency phase-shifted waves obtained for a plasmonic reference (black) $V_{DC}$ = 2.0 V, $I_t$ = 70 pA, $V_{AC}$ = 100 mV, integration time 180 s; neutral exciton X (red) $V_{DC}$ = -2.2 V, $I_t$ = 70 pA, $V_{AC}$ = 100 mV, integration time 600 s; and trion $X^+$(purple) $V_{DC}$ = -3.2 V, $I_t$ = 120 pA, $V_{AC}$ = 100 mV, integration time 600 s. X and $X^+$ waves were taken above the ZnPc lobe. The lines mark the phase shift between excited states and the reference. (**C**) STML spectra obtained on a lobe and a point near the center of ZnPc. $V_{DC}$ = -3.0 V, $I_t$ = 100 pA, integration time 180 s. The locations of charge injection are marked in the insets. (**D**) Exciton (red) and trion (purple) intensity dependence on the applied $V_{DC}$ at $I_t$ = 40 pA.

Electroluminescence intensities of X and $X^+$ strongly depend on precise location of the charge injection in the molecule (Fig. 1C) and on the $V_{DC}$ (Fig. 1D). On 3 layers of NaCl on Ag(111), X and $X^+$ generation is activated at negative $V_{DC}$ with thresholds of -1.9 V and -2.6 V, respectively. The exact value of the thresholds may vary as much as 0.2 eV depending on exact tip condition. Within the picocavity, the $X^+$ threshold voltage ($V_{th}$) represents a sizable electric field which triggers transient positive charging in the neutral excited molecule. *(9,13,26)* We find that a hole injection into the peripheral aromatic rings boosts the emission from the $X^+$ (Fig. 1C), suggesting that the two excitations are induced by two separate charge transport channels. *(27)*

Bias-dependent hyperspectral mapping with p- and s-wave probes using CO-functionalized and metallic apexes (Fig. 2A) reveals a remarkable spatial anticorrelation between the charge injection locations inducing X and $X^+$ emission. X intensity appears concentrated at the porphyrazine macrocycle, extending toward the peripheral benzene rings at higher magnitudes of $V_{DC}$. All $X^+$ maps observed at $V_{DC} < V_{th}$ show a 12-lobe corral-like pattern with the highest intensity distributed around the periphery of the molecule. $X^+$ map taken for comparison using a metal apex displays the intensity clearly concentrated at the peripheral benzenes, confirming that the charge injection into them strongly promotes the transient state leading to the formation of trions. At constant height, X and X$^+$ lines red-shift a few nm, which is apparent on a spectral map from a diagonal ZnPc cross-section and maps for each exciton at particular wavelengths (see Fig. 2B-D). This may be explained in terms of photonic Lamb shift by strong coupling between the optical excitations of ZnPc and the highly confined plasmons and has been observed by tip-enhanced plasmon absorption and photoluminescence. *(28-30)*

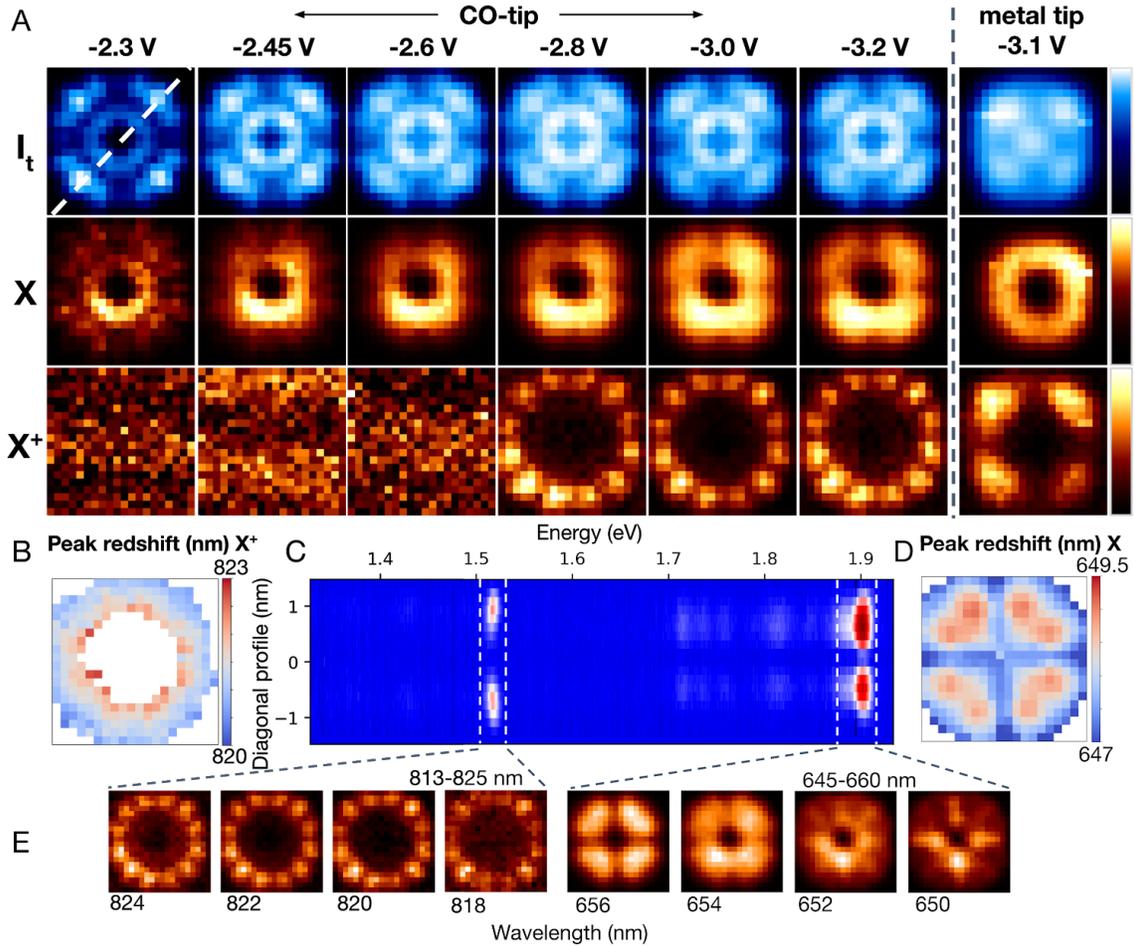

**Figure 2. Mapping of the exciton and trion.** (**A**) Constant-height tunneling images ($I_t$) and simultaneously measured electroluminescence photon maps of the neutral exciton (X) and trion (X$^+$) of a single ZnPc molecule, using a CO-functionalized- and a metallic tip. The X and X$^+$ maps represent the total photon intensities in the 645-660 nm and 813-828 nm, respectively. (**B**) Red-shift map of the X$^+$ line. (**C**) spectral cross-section map measured across a diagonal denoted in (A) (white dashed line). (**D**) Redshift map of X line. (**E**) Photon maps at selected wavelengths showing variation of intensity as a function of charge injection positions. (B), (D), (E) were taken with a CO-functionalized tip at $V_{DC}$ = -3.0 V. All images are 1.8 x 1.8 nm$^2$.

The strong bias dependence of the X and X$^+$ emission motivated us to study the dependence of the dynamics on the tunnelling conditions. Series of RF-PS measurements reveal the tunability of the X and X$^+$ effective lifetimes depending on the $V_{DC}$ and on the tunneling gap size given by the relative tip-molecule distance ($Z_{rel}$), and hence on the tunnelling current ($I_t$). X and X$^+$ lifetimes (shown in Fig. 3A) rise monotonically with increasing $Z_{rel}$ (and with decreasing $I_t$), showing that closing of the picocavity significantly promotes quenching of the excitations. The trends agree with previous measurements on similar systems. *(14,16)* Varying $V_{DC}$ and maintaining the current in the 50 - 130 pA range, we observe a X lifetime drop from values well above 700 ps to as low as 200 ps upon crossing the $V_{th}$ (see Fig. 3B). This drop in the lifetime is also reflected in the X intensity. In the intensity *vs.* $V_{DC}$ measurement for X in Fig. 1D we note a superlinear rise of the photon rate for voltages beyond $V_{th}$, implying that around this value the average interval between photons reduces. At -3.2 V, X lifetime recovers to around 450 ps. The trion ($X^+$) lifetime follows a tendency similar to the neutral exciton (X), growing from below the experimental resolution limit at $V_{th}$ to near 600 ps at $V_{DC}$ < -3.0 V (Figs.3B). Such similar trends for X and $X^+$ dynamics as a function of voltage indicate that the generation mechanisms and dynamics of both excitations are closely related. The dependence on $V_{DC}$ and $Z_{rel}$ of the lifetimes highlights the crucial role of the picocavity in reducing the observed radiative lifetimes of the chromophore. The neutral exciton lifetime of ZnPc in solution has been reported to be 4.73 ns. *(31)* In our experiments we find an increase of the measured lifetimes up to 4.8 ns after increasing the number of NaCl layers and using Au(111) as a substrate. *(23)*

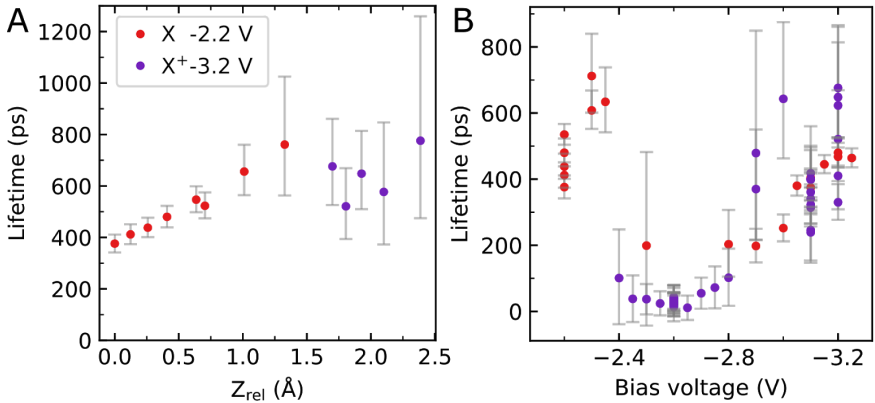

**Figure 3. Dependence of lifetimes on picocavity size and bias voltage.** (**A**) Lifetime of neutral exciton (red) and trion (purple) as a function of tip-molecule relative distance ($Z_{rel}$). (**B**) Lifetime of neutral exciton (red) and trion (purple) as a function of applied $V_{DC}$. A 200 MHz harmonic signal with 100 mV amplitude is used for measuring the phase shift and determination of the lifetimes. Range of $I_t$ was 50 to 130 pA.

It remains to be verified that the X and $X^+$ are not only spatially, but also temporally anticorrelated. This behavior is anticipated, as both excitons are localized within the same molecule and they cannot coexist at the same moment. We take the advantage of the strong nonlinearity of $X^+$ with bias, and apply modulation around $V_{th}$ in order to generate anharmonic wave on the output, looking for hallmarks of an interdependence between the X and $X^+$ formation. By using a $V_{DC}$ value of -2.4 V and a RF amplitude of 250 mV the $X^+$ is excited only in the first half of the sinusoidal driving while X can be excited during the entire cycle (Fig. 4A). The purpose is to reach conditions at which the time-resolved probability of X formation is dented by the sharply growing probability of charging the molecule and forming $X^+$. In Fig. 4B it is evident that the resulting X wave is indeed distorted in the upper portion when $X^+$ intensity rises, bringing the ultimate evidence that the X and $X^+$ represent two mutually exclusive states of the molecule.

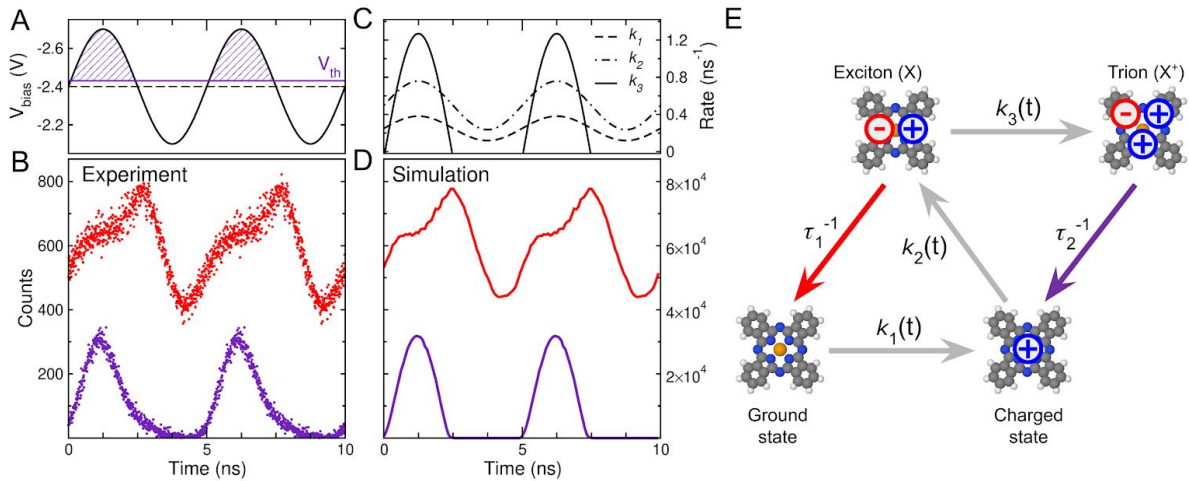

**Figure 4. Temporal anticorrelation of the exciton and trion and simulation.** (**A**) Bias voltage is harmonically modulated across the trion threshold at 200 MHz, $V_{DC}$ = -2.4 V and $V_{AC}$ = 250 mV, $I_t$ = 70 pA. (**B**) RF-PS waves recorded for the neutral exciton (red) and trion (purple). Dark counts have been subtracted from the experimental waves. (**C**) Rates of the time-dependent transitions used in the simulation. (**D**) Simulated RF-PS waves for the neutral exciton (red) and trion (purple). (**E**) Scheme of the 4-state model with transitions used for simulation of the exciton-trion dynamics in the ZnPc molecule in (D).

Simulation of the internal charging and excitation dynamics is possible with a causally deterministic state model working with a minimized number of transitions and states, schematically shown in Fig. 4E. We define four states: the ground state ($ZnPc^0$), positively charged molecule ($ZnPc^+$), and the excited states X and $X^+$. The types of transitions allowed in the framework of this model are i) hole injections, ii) electron captures and iii) decays of the excited states. Transitions are simulated either as homogeneous (time-independent) or inhomogeneous (time-dependent) Poissonian processes. The decay rates of the X and $X^+$ are fixed according to the values of the lifetimes experimentally obtained near $V_{th}$ (1/750 $ps^{-1}$ and 1/50 $ps^{-1}$, respectively). The rates of the hole injection probabilities from $ZnPc^0$ to $ZnPc^+$ are modulated with a sine and from X to $X^+$ with a half-sine (Fig. 4C). *(23)* Excitons are achieved via electron capture by $ZnPc^+$ at a rate that is sine modulated. The model with optimized rates presented in Fig. 4D and the measured histograms in Fig. 4B reach a very good agreement comparing relative count intensities, phases and shapes for both X and $X^+$ waves, without the necessity to include additional transitions or states.

The model describes the internal mechanism of the transient ZnPc charging under the external potential and the exciton-trion dynamics. It validates the explanation of the experimentally observed maps of excitons by increase of the hole trapping probability of the ZnPc above the molecular lobes. Importantly, it shows that trions are primarily generated through an exciton-to-trion conversion mechanism upon hole trapping, analogous to the one observed for optical excitations in strained 2D materials. *(32)* The trapping efficiency is greatly enhanced if charge is injected into the peripheral aromatic rings of the molecule (Fig. 2A), where it triggers more easily the renormalization of the electronic structure to form a trion bound state. It remains to be clarified how the lifetime dependence on bias of the two excitons are connected, especially around the $V_{th}$. The RF phase fluorometry method implemented in a scanning probe microscope is suitable for determination of exciton dynamics at the nanoscale in other molecular adsorbates on surfaces and low dimensional systems. In addition, the fast trion lifetimes measured here opens new avenues to use single molecules as single photon electrooptical transducers in the GHz range .

**Acknowledgments:**

The authors are grateful to our colleague Dr. Lukáš Ondič for sharing the optical instrumentation. **Funding:** M. Š. and J. D. acknowledge the Czech grant agency funding no. 20-18741S and the Charles University Grant Agency project no. 910120. P. M. thanks the ERC Synergy Program (grant no. ERC-2013-SYG-610256, Nanocosmos) for financial support and the "Comunidad de Madrid" for its support to the FotoArt-CM Project (S2018/NMT-4367) through the Program of R&D activities between research groups in Technologies 2013, co-financed by European Structural Funds.

# Supplementary Information for

# Exciton-trion dynamics of a single molecule in a radio-frequency cavity

**Contents:**

**Materials and methods**

The experiments were performed in ultrahigh vacuum (base pressure of 5x10-11 mbar), low temperature (7 K) conditions in a Createc STM/AFM instrument extended with an optical path allowing to focus and redirect photons originating from the tunnel junction into an CCD-spectrograph and an avalanche photon detector (APD). Light is first collimated by an achromatic lens in the cryostat with a maximum theoretical collection of 3,64% of the full sphere, and guided through the two cryostat shields (sapphire windows) and out of the vacuum through a fused silica viewport. Outside the vacuum chamber, the beam is split by a beam splitter 90:10 and the 90% branch is filtered by a bandpass filter and coupled to a 200 μm optical fiber, achromatically coupled to the active area of Perkin-Elmer SPAD (SPCM-AQR-15, jitter time

~250 ps). The 10% branch is refocused to a fiber bunch, formed into a slit termination inserted into Shamrock 163i spectrograph with DU401-BV CCD camera, with spectral detection range of 400-900 nm, and resolution ~ 2.5 nm FWHM.

As scanning probes, 30 kHz Q-plus sensors were used with PtIr-tips sharpened ex-situ using a focused Xe ion beam (Tescan FERA3). The resulting quality factors of the sensors at 7 K were >20000. The probes were ultimately formed by means of controlled nanoindentations into the clean substrate to coat them by Ag or Au and to optimize their plasmonic response at the desired optical frequencies. Ag(111) and Au(111) substrates were prepared by conventional sputtering and annealing cycles prior to thermal NaCl deposition (at 610°C, 5 min), which was done with the substrates held at 100°C to form bi- tri- and tetralayer islands of NaCl. Subsequently the samples were transferred into the cryostat, cooled < 10 K and exposed to a flux of ZnPc molecules, which were evaporated at 425°C, 3 min from a Ta crucible. All optical spectra were collected in constant-current mode with the feedback loop switched on; the photon maps were collected in the constant height height mode with the feedback loop switched off.

Radio frequency produced by an arbitrary wave generator (Keysight, 81160A) was added to the DC tunneling bias voltage generated from the SPM control electronics (Nanonis GmbH) using a bias-tee (Pasternack, PE1608) and connected to the bias wiring of the microscope on the air side feedthrough. The transmission of the wiring was calibrated at 200 MHz using broadening of a plasmon high-energy cutoff at 1.8 V on a clean substrate. The amplitudes reaching the picocavity were varying between 50 and 400 mV peak-to-peak. Synchronization signal from the arbitrary wave generator at 20 MHz (ten cycles) is used as the start trigger for photon arrival time tagging and the APD signal corresponding to the first incoming photon is employed as the stop trigger. Therefore the range of the time frame was 50 ns. The time tagging and histogramming of the photon arrival times is performed by dedicated counting electronics and software (Picoquant, PicoHarp 300). Typical bin size was 64 ps For filtering of the X and X+ exciton emission we used hard-coated 25 nm bandpass filters (Edmund optics) with center wavelengths 650 nm (for measurements of X lifetime) and 825 nm (for X+). Histograms are accumulated for approximately 3-18 mins, depending on the strength of the modulation and the resulting signal-to-noise ratio in the waves.

**Relation of the phase shift and delay**

The base concept of the RF-PS method is a harmonically-modulated electrical driving of the system with frequency $f$ and detection of the variation in photon output.[33] The electroluminescent response, i.e. the decay rate of the excitons formed as a result of the electric field, will follow this driving, however its finite lifetime will cause a phase shift $\Delta\varphi$ with respect to the phase of the driving signal. Given that the probability of finding the system in the excited state at time $t$ after excitation decreases *monoexponentially* following the expression $exp(-t/\tau)$, where $\tau$ is the intrinsic lifetime characteristic of the exciton, it can be shown that the relation of $\Delta\varphi$ and $\tau$ is

$$tan(\Delta\varphi) = 2\pi f\tau \qquad (1)$$

This implies that for optimum detection of the lifetime, the corresponding $\Delta\varphi$ has to be well below $\pi/2$ (to avoid divergence of the tan function and excessive attenuation of the output) but at the same time sufficiently high to allow precise-enough detection. This can be ensured by setting an adequate value of $f$. From the simulated $\Delta\varphi$ dependence on $f$ and $\tau$ in Fig.S1, we can see that for example $f$ = 200 MHz is suitable for lifetime range of 70 - 1200 ps and $f$ = 50 for the range of 0.250 - 5 ns.

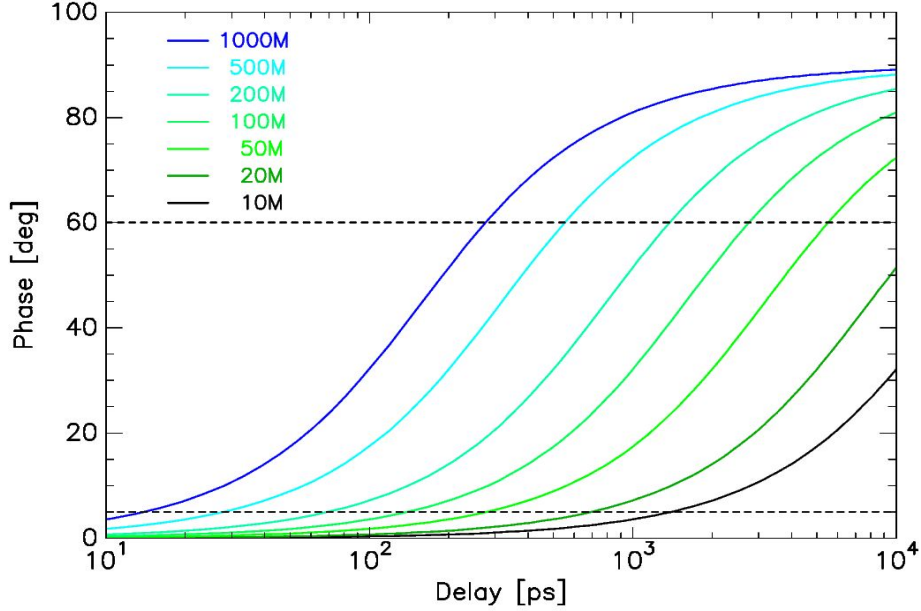

**Fig. S1**: Simulation of phase shifts dependence on lifetime-related electroluminescent response delay of a system for different voltage modulation frequencies.

**Determination of the phase shift, amplitude and error bars of the RF-PS waves**

The periodic photon arrival time histograms (waves) are represented as N couples of discrete bin time tag and photon counts $(t_i, c_i)$. The absolute phase of each wave has been determined as

$$\varphi = \tan^{-1}\left( \sum_{i=1,N} c_i \cdot \sin(2\pi f \cdot t_i) \, / \, \sum_{i=1,N} c_i \cdot \cos(2\pi f \cdot t_i) \right) \qquad (2)$$

and the amplitude $A$ as

$$A^2 = \left[ \sum_{i=1,N} c_i \cdot \sin(2\pi f \cdot t_i) \right]^2 + \left[ \sum_{i=1,N} c_i \cdot \cos(2\pi f \cdot t_i) \right]^2 \qquad (3)$$

For representation in the graphs, the measured waves (ten periods) have been folded in a single period by performing a modulo operation on the $t_i$ values, obtaining $(t_i \bmod 1/f, c_i)$. We should note that the $\varphi$ changes by $\pi$ upon changing the offset bias polarity at the tunneling junction.

If we assume the noise in the data is white and neglect any quantization noise in the waves, the variance of the phase and amplitude determination scales with by the standard deviation $\sigma$ and N as $\sigma^2(2/N)$. Value of $\sigma$ is calculated as the difference of the wave and its idealized form

$$\sigma^2 = (N-1)^{-1} \cdot \Sigma_{i=1,N} [ c_i - <c_i> - A.\sin(2\pi f.t_i + \varphi) ]^2 \qquad (4)$$

Higher and lower error bars of the $\tau$ are then obtained in a standard way combining the Eqs. 1 and 4 and considering that $\Delta\varphi = \varphi - \varphi_{REF}$, where $\varphi_{REF}$ is the reference phase measured using plasmonic emission. It is important to note that the effect of detector jitter leads to temporal smearing of the signal and decrease of $A$, however these effects are very mild if jitter is well below the $1/f$, which is the case in our experiments. Therefore $\sigma$ can be significantly lower than the jitter.

**Measurement of the reference phase**

Impedance of the instrumental wiring causes a significant change of the modulation phase and amplitude reaching the junction with respect to the input from the radio frequency generator. The reference phase at the junction can be measured with a high precision using the plasmonic emission from the picocavity formed between tip and substrate. The same setup and tip conditions are used right after for measuring the phase-shift (and lifetime) of the modulated light of excitonic origin. Plasmon lifetimes are typically in the fs range, resulting in a negligible shift in comparison to excitons, making it therefore an ideal calibration measurement of the absolute radio-frequency phase reaching the cavity. Nevertheless, the speed of light propagation through the optical system (mostly through the fibers) is influenced by wavelength-dependent refraction index, which can result in a difference of up to 150 ps between the time-of-flight of photons at 650 and 825 nm using a 2 m fiber leading to the APD. The reference phases for the $f = 200$ MHz used in the experiments were therefore determined for X and $X^+$ wavelengths independently, taking advantage of the broad nature of the plasmon spectrum and using exclusively the plasmonic photons filtered by the same filters that have been used for measurements of the neutral exciton and trion histograms. The typical error in determining the reference phase, evaluated according to the scheme described above, was below 15 ps.

**Implementation of the state model simulation**

The kinetic model is based on discrete states between which the system changes stochastically with defined rates $k_a$ ($a$ being the index of a particular allowed transition). The decay rates from the X and X$^+$ states are kept fixed in time, as they characterize the intrinsic lifetimes of the excitations which are assumed to stay constant in the voltage ranges used in the experiments. However all other transition rates which represent the charging probability rates (i.e. hole injections and electron captures) are simulated by periodic functions of time, i.e. $k_a(t)$ which account for the varying tunnelling probability at different bias voltages.

The numerical simulation is performed as a sequence of homogeneous and inhomogeneous Poissonian processes, with the aim to obtain a series of individual event times $t_n$, and its subsets $T^a_j$ for each allowed transition. For a large enough total number of transitions $j$ (typically on the order of $10^6$), the distributions of the transition events $T^a_j$ (which correspond to the X, X$^+$ emission events) in a single period can be directly compared to the experimental data.

For an increased efficiency of the simulation, we choose to simulate the time intervals it takes the system to change from one state to another by any of the allowed transitions, For a homogeneous Poissonian process they can be obtained as

$$\Delta T^a_j = -\ln(r_n) / \Sigma_x k_x \qquad (5)$$

where $r_n$ is a random number with uniform distribution in the range (0,1) generated for each transition[34] and the sum runs over all possible states that can follow the current state. The moment of transition $T^a_j$ will be

$$T^a_j = t_n = t_{n-1} + \Delta T^a_j. \qquad (6)$$

The probability that a transition will undergo a particular allowed transition is

$$P^a_j = k_a / \Sigma_x k_{x'} \qquad (7)$$

which is simulated by a Monte Carlo approach in a straightforward manner using a second uniformly distributed random variable $q_n$.

In the case of time-dependent rates $k_a(t)$, the simulation of the $\Delta T^a_j$, a general analytical solution is not feasible, however it can be estimated numerically with a very good precision using again the random variable $r_n$. The $\Delta T^a_j$ can be obtained by integrating the total time-variable rate since the last event at time $t_{n-1}$, until the value of the randomized value of $r_n$ is reached:[34]

$$r_n = \int_0^{\Delta T} \Sigma_x k_x(t_{n-1} + s)\, ds \qquad (8),$$

$1/f$ periodicity of $k_a(t)$ gives the possibility to speed up the algorithm by precalculating the integral and creating a lookup function to quickly find $\Delta T^a_j$ for each pair of $r_n$ and $t_{n-1}$ mod $1/f$. Our discrete-step integration is done with a step $ds = 1$ ps, up to $s = 5.10^4$ ps. For such $r_n$ and $t_{n-1}$ mod $1/f$ that reach the upper bound of $s$, the $\Delta T^a_j$ is laid equal to the value at the maximum $s$.

The probability of transition $P^a_j$ follows the Eqn.7, however with the time-variable rates at time $t_n$

$$P^a_j = k_a(t_n) / \Sigma_x k_x(t_n) \qquad (9)$$

In the four-state model, we included the transitions and the time-variable rates listed in Table.S1.

| Name | Transition | Movement of charge | Rate |
|---|---|---|---|
| $k_1$ | ZnPc$^0$ to ZnPc$^+$ | h$^+$ injection | $\boldsymbol{\kappa_1}.[1 + \boldsymbol{\alpha}.\sin(2\pi f.t)]$ |
| $k_2$ | ZnPc$^+$ to X | e$^-$ capture | $\boldsymbol{\kappa_2}.[1 + \boldsymbol{\alpha}.\sin(2\pi f.t)]$ |
| $k_3$ | X to X$^+$ | h$^+$ injection | $\sin(2\pi f.t) > \boldsymbol{L}$: $\boldsymbol{\kappa_3}.[\sin(2\pi f.t)-\boldsymbol{L}]^{\boldsymbol{w}}$ <br> $\sin(2\pi f.t) \leq \boldsymbol{L}$: 0 |
| $\tau_1^{-1}$ | X to ZnPc$^0$ | recombination | $1/\tau_1$ |
| $\tau_2^{-1}$ | X$^+$ to ZnPc$^+$ | recombination | $1/\tau_2$ |

**Table S1**: Transitions allowed in the minimalistic four-state model and their corresponding time-dependent rates. The parameters of the model are emphasized in bold.

The $k_1$ rate of hole injection into the molecule is a product of the base rate $\kappa_1$ sinusoidally modulated with amplitude $\alpha$, set to 0.5. The rate of electron capture into ZnPc$^+$ $k_2$ is analogous to $k_1$, however with its own rate $\kappa_2$. The hole injection probability into exciton $k_3$ is modulated by a sine above a threshold L, raised to the power $w$ and multiplied with their respective base rates $\kappa_3$. The $w$ and $L$ have the purpose of accounting for the nonlinear rise of the X$^+$ above the threshold, and we set them to $w$ = 1.4, $L$ = 0.1. Lifetimes $\tau_1$ and $\tau_2$ defining the X and X$^+$ decay rates are kept at 750 and 50 ps, respectively, which are the characteristic values measured around V=V$_{th}$. The $\kappa_1$, $\kappa_2$, $\kappa_3$ are optimized by iterative least-squares approach to the best match of the simulated and experimental X and X$^+$ decay histograms. The experimental histogram for the X$^+$ was multiplied by a factor accounting for different relative sensitivities of the entire optical setup to the wavelengths of X and X$^+$.

**Lifetime measurements on different locations of a molecular trimer**

In order to test the role of the injection position on the determination of the excitonic lifetime we have constructed a ZnPc trimer and measured the neutral exciton lifetime across the structure by means of RF-PS. Porphyrin chains have been shown to behave as single photon emitters due to dipole-dipole coupling of the individual monomers, which results in superluminescence of the coupled emitters.[8] The lifetimes of the excitons in the chains as measured by Hanbury Brown-Twiss interferometry, however, remain close to the one observed for single molecules (see Ref. [25] for details). In Fig. S2 we show that lifetime values measured along the long axis of a trimer remain constant within the error of the measurement. The small modulation observed may be indicative of the different charge injection efficiencies due to the spatial modulation of the molecular orbitals, and highlights the convenience of using RF-PS measurements for studying the photophysics of coupled emitters.

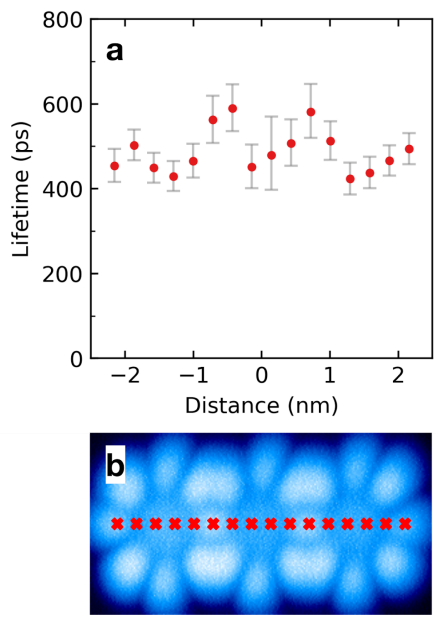

**Fig. S2**: a) **Lifetime as a function of the tip position above the ZnPc trimer, measured along its long axis. Frequency was 200 MHz, $V_{DC}$ = -2.2 V and $V_{AC}$ = 100 mV, $I_t$ = 100 pA.** b) **Constant current STM image ($V_{DC}$ = -2.2 V, $I_t$ = 2 pA). Red crosses denote the locations of the lifetime measurements drawn in (a)**.

**Lifetime measurements of ZnPc on 4 ML-NaCl/Au(111)**

We performed lifetime measurements of ZnPc on 4 ML-NaCl/Au(111) system in which neutral exciton appears also at positive bias voltages where the trion is absent[12]. Very sharp bias onset of the measured exciton intensity and high quantum yield allowed us to measure lifetimes up to 4.8 ns at 10 pA current. The tip is closer to the molecule at same current due to the reduced conductivity of 4 ML of NaCl than on 3 ML. The resulting lifetimes are approximately 5 times longer than for ZnPc on 3 ML-NaCl/Ag(111) and comparable to values reported for ZnPc in solutions[31]. This showcases the usefulness of the method for measurements of longer lifetimes.

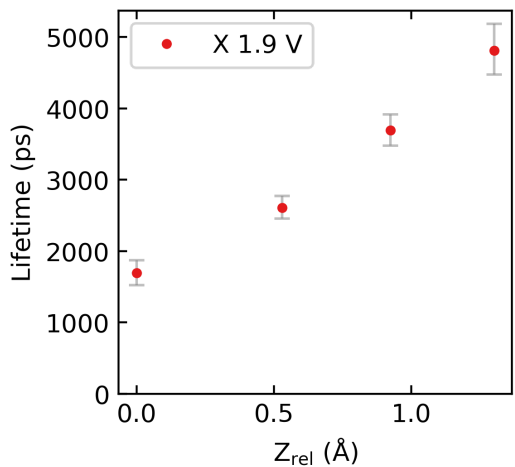

**Fig. S3**: Dependence of lifetime of neutral exciton in ZnPc/4 ML-NaCl/Au(111) on picocavity size measured at positive bias voltage of 1.9 V. A 50 MHz harmonic signal with 100 mV amplitude is used for measuring the phase shift and determination of the lifetimes. Range of $I_t$ was 10 to 60 pA.

**Redshift maps measured with a metallic tip**

To complement the redshift map presented in Fig. 2D obtained with a CO-tip we also derive an analogous picture using the data taken with a metallic tip. The map in fig. S3 is showing the spectral shift of the X peak as a function of the tip position (and hence primary charge carrier injection location). It closely resembles the data recently reported for photonic Lamb shift in photoluminescence maps of ZnPc.[30] The observed shift can be explained by means of plasmon–exciton interactions at the submolecular level and proves that in the electroluminescent process the role of the picocavity plays a crucial role and can be used for fine tuning of the emission energy. Indeed, comparison of Fig. S3 with Fig. 2D demonstrates how a subtle modification (by picking up a CO molecule with the tip) can alter the spatial character of the photonic Lamb shifts; redshift map with CO-functionalized tip presents eight lobes whereas with metallic tip we observed only four. Future simulations taking into account the presence of the CO probe particles in the picocavities may help to elucidate the photophysical origin of these apparent differences.

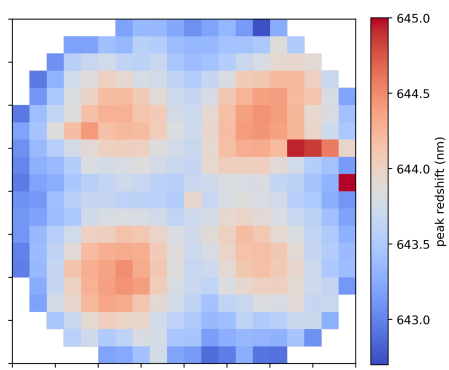

**Fig. S4:** Peak redshift map of neutral exciton on a single ZnPc measured with a metallic tip at $V_{DC}$ = -3.1 V. The redshifts are derived from the dataset presented in Fig. 2A.